\documentclass[twocolumn,preprintnumbers,superscriptaddress,nofootinbib,showkeys,aps,prd,floatfix]{revtex4}

\usepackage[utf8]{inputenc}

\usepackage{footmisc,enumerate}
\usepackage{multirow}
\usepackage{subfigure}
\usepackage{amsmath}
\usepackage{graphicx}
\usepackage{placeins}
\usepackage{hyperref}
\hypersetup{colorlinks=true, citecolor=blue, urlcolor=blue, linkcolor=blue}
\usepackage{tabulary}
\newcolumntype{K}[1]{>{\centering\arraybackslash}p{#1}}

\renewcommand{\vec}[1]{{\bf{#1}}}

\newcommand{\dst}{$|{\cal{M}}_\text{d6}|^2$}
\newcommand{\dsix}{dimension-6~}
\graphicspath{{./results/}}


\begin{document}

\preprint{IPPP/18/68. This article is registered under preprint number: arXiv:1808.06577}

\title{Angles on CP-violation in Higgs boson interactions}

\begin{abstract}
CP-violation in the Higgs sector remains a possible source of the baryon asymmetry of the universe.  Recent  differential measurements of signed angular distributions in Higgs boson production provide a general experimental probe of the CP structure of Higgs boson interactions.  We interpret these measurements using the Standard Model Effective Field Theory and show that they do not distinguish the various CP-violating operators that couple the Higgs and gauge fields.  However, the constraints can be sharpened by measuring additional CP-sensitive observables and exploiting phase-space-dependent effects.  Using these observables, we demonstrate that perturbatively meaningful constraints on CP-violating operators can be obtained at the LHC with luminosities of ${\cal{O}}$(100/fb). Our results provide a roadmap to a global Higgs boson coupling analysis that includes CP-violating effects.
\end{abstract}

\author{Florian U.\ Bernlochner} \email{florian.bernlochner@kit.edu}
\affiliation{Institut f\"ur Experimentelle Teilchenphysik (ETP), KIT, 76131 Karlsruhe, Germany\\[0.1cm]}
\author{Christoph Englert} \email{christoph.englert@glasgow.ac.uk}
\affiliation{SUPA, School of Physics \& Astronomy, University of Glasgow, Glasgow G12 8QQ, UK\\[0.1cm]}
\author{Chris Hays} \email{chris.hays@physics.ox.ac.uk}
\affiliation{Department of Physics, Oxford University, Oxford, OX1 3RH, UK\\[0.1cm]}
\author{Kristin Lohwasser} \email{kristin.lohwasser@cern.ch}
\affiliation{Department of Physics \& Astronomy, Sheffield University, Sheffield S3 7RH, UK\\[0.1cm]}
\author{Hannes Mildner} \email{hannes.mildner@cern.ch}
\affiliation{Department of Physics \& Astronomy, Sheffield University, Sheffield S3 7RH, UK\\[0.1cm]}
\author{Andrew Pilkington} \email{andrew.pilkington@manchester.ac.uk}
\affiliation{School of Physics \& Astronomy, University of Manchester, Manchester M13 9PL, UK\\[0.1cm]}
\author{Darren D.\ Price} \email{darren.price@manchester.ac.uk}
\affiliation{School of Physics \& Astronomy, University of Manchester, Manchester M13 9PL, UK\\[0.1cm]}
\author{Michael Spannowsky} \email{michael.spannowsky@durham.ac.uk}
\affiliation{Institute of Particle Physics Phenomenology, University of Durham, Durham DH1 3LE, UK\\[0.1cm]}


\maketitle

\section{Introduction}
\label{sec:intro}

The matter-antimatter asymmetry of the Universe provides one of the primary motivations to study extensions of the Standard Model of Particle Physics (SM). Various ideas have been proposed \cite{Fukugita:1986hr,Affleck:1984fy} that satisfy the Sakharov conditions \cite{Sakharov:1967dj} to generate a net surplus of matter over anti-matter throughout the Universe. One of the most popular and thoroughly studied approaches is electroweak baryogenesis \cite{Kuzmin:1985mm, Shaposhnikov:1987tw,Nelson:1991ab, Morrissey:2012db}, which requires C- and CP- violating interactions to be present during a phase of expanding regions of non-trivial vacuum expectation value, i.e. a strong first-order phase transition. 
Although the SM provides a source for baryon and lepton number violation through sphaleron transitions \cite{Klinkhamer:1984di, Kuzmin:1985mm, Ringwald:1989ee}, it falls short of a complete explanation of the matter-antimatter asymmetry on two accounts: the CP-violating phase in the weak interactions of quarks is not large enough to explain the observed asymmetry and the electroweak phase transition is not a strong first-order phase transition.

In extensions of the SM, the required CP violation is often obtained by introducing complex phases into the scalar sector \cite{Branco:2011iw} or into Higgs-fermion interactions~\cite{Farrar:1993sp, Bernreuther:2002uj}. 
Precision studies of the Higgs boson interactions, in particular CP-violating interactions, can therefore provide a window into the dynamics of the early Universe and can help us to unravel the mechanism underlying the matter-antimatter asymmetry.\footnote{Modifications to the Higgs boson self coupling can induce a strong first-order electroweak phase transition~\cite{Grojean:2004xa,Kanemura:2004ch,Kanemura:2005cj,Ham:2005ej}. The measurement of the Higgs boson self coupling therefore complements searches for CP violation in the Higgs sector to explain the matter-antimatter asymmetry.}
The current constraints on the Higgs boson interactions provided by the ATLAS and CMS experiments at the Large Hadron Collider (LHC) still allow coupling deviations of 10\% (or larger) from the SM predictions~\cite{Khachatryan:2016vau,Mariotti:2016owy,Butter:2016cvz,Englert:2015hrx,Englert:2017aqb,Ellis:2018gqa,deBlas:2018tjm}. As such, they are too loosely constrained to provide a fine-grained picture of electroweak symmetry breaking.  Analysing the detailed properties of the Higgs boson therefore remains at the heart of the LHC research programme.

In this Letter, we present combined constraints on CP-odd operators in the Higgs sector using only measurements sensitive to interference between these operators and those of the SM (the physics potential of such an analysis was discussed recently in Ref.~\cite{Brehmer:2017lrt}).  We start with ATLAS measurements of model-independent differential cross sections in the $h\rightarrow \gamma\gamma$ and $h\rightarrow ZZ^* \rightarrow 4\ell$ decay channels~\cite{Aaboud:2018xdt,Aaboud:2017oem}. Model-independent data are crucial in avoiding unnecessary assumptions about the nature of specific Higgs boson couplings. The data are reinterpreted using the dimension-6 Standard Model Effective Field Theory (SMEFT)~\cite{Grzadkowski:2010es}. The EFT is linearised to ensure that constraints on CP-violating operators in the EFT are driven entirely by CP-sensitive observables, whereas kinematic information such as transverse momentum distributions are instead used to constrain CP-even operators.

The ATLAS experiment has performed two CP-sensitive differential measurements of the signed azimuthal angle between the hadronic jets in $h+2~{\rm jet}$ events.  We use these measurements to calculate a combined asymmetry in this angle of $0.3 \pm 0.2$.  Interpreting the results in the SMEFT, we find that the current data cannot distinguish between different sources of CP violation, with three blind directions when one considers the four CP-odd operators that cause anomalous Higgs boson interactions with weak bosons or gluons. We then demonstrate how the blind directions in the CP-odd coupling space can be removed using observables that can already be measured with the existing LHC datasets. Building on these insights, we provide projections for the upcoming LHC Run-3 and the high-luminosity LHC (HL-LHC), where the available dataset will increase by factors of 10 and 100, respectively.

The Letter is organised as follows. We motivate the linearised dimension-6 effective field theory in Sec.~\ref{sec:basics}. Section~\ref{sec:fit} provides an overview of technical aspects of our analysis. The constraints on EFT operators obtained by fits to published model-independent data are presented in Sec.~\ref{sec:results_curr}. 
In Sec.~\ref{sec:results_proj} we propose new measurements be made
and show their expected impact on constraining the different sources of CP violation in the Higgs sector. 
Finally, we conclude in Sec.~\ref{sec:conc}.

\section{Theoretical framework}
\label{sec:basics}

New CP-violating effects in the Higgs boson's interactions with gluons or weak bosons can be introduced through a minimal set 
of CP-odd \dsix operators~\cite{Grzadkowski:2010es}:
\begin{subequations}
\label{eq:ops}
\begin{alignat}{5}
\label{eq:opodd}
O_{H\tilde{G}}&=H^\dagger H G^{a\mu\nu} \tilde G^{a}_{\mu\nu}\,,\\
O_{H\tilde{W}}&=H^\dagger H W^{a\mu\nu} \tilde W^{a}_{\mu\nu}\,,\\
O_{H\tilde{B}}&=H^\dagger H B^{\mu\nu} \tilde B_{\mu\nu}\,,\\
O_{H\tilde{W}B}&=H^\dagger \tau^a H  B_{\mu\nu} \tilde {W}^{a \mu\nu}\,,
\end{alignat}
\end{subequations}
where $H$ is the Higgs doublet and $G,W,B$ are the $SU(3)\times SU(2)\times U(1)$ field strength tensors. The $\tau^a$ are the SU(2) generators. Fields with a tilde 
are the dual tensors, e.g. $\tilde{G}^{a}_{\mu \nu}=\varepsilon^{abc} G^{bc}_{\mu \nu}/2$. 

These operators could originate from complex phases in the interactions between the Higgs boson and heavy fermions, whose masses 
are far above the electroweak scale.  Additional complex phases in the SM Yukawa sector would be another source of CP-violation, 
e.g. in the $t\bar{t}h$ interaction~\cite{Ellis:2013yxa,Casolino:2015cza,Buckley:2015vsa,Goncalves:2018agy}.  Any kinematic 
effect from this interaction would be degenerate with $O_{H\tilde G}$ in gluon-fusion production as long as the $m_t$ threshold 
is not resolved kinematically, which does not happen for our choice of measurements.  An associated blind direction is therefore 
implied in our constraints.

The operators of Eq.~\eqref{eq:ops} are well-motivated candidate interactions for our analysis.  They are closed under RGE 
flow~\cite{Jenkins:2013zja,Jenkins:2013wua,Alonso:2013hga,Grojean:2013kd,Elias-Miro:2013gya}, allowing well-defined constraints. 
Furthermore, the small number of operators can be probed with a few differential distributions.

For completeness, analogous CP-even deformations to the SM are also introduced ($O_{HG},O_{HW},O_{HB},O_{HWB}$). 
The effective Lagrangian is then defined as
\begin{equation}
{\cal{L}} = {\cal{L}}_{\text{SM}} + \sum_i {c_i\over \Lambda^2} O_i
\end{equation}
where the sum runs over the CP-even and CP-odd operators. This allows us to split the amplitude into an SM part, 
${\cal{M}}_\text{SM}$, and a genuine Beyond the Standard Model (BSM) \dsix part, ${\cal{M}}_\text{d6}$. 
Including all \dsix effects yields
\begin{equation}
\label{eq:ampexp}
|{\cal{M}}|^2 = |{\cal{M}}_\text{SM}|^2 + 2\text{Re}\left( {\cal{M}}_\text{SM}^\star {\cal{M}}_\text{d6}\right) + 
{\cal{O}}(\Lambda^{-4}).
\end{equation}
The integration over interference terms (proportional to $1/\Lambda^2$) vanishes when only 
CP-odd EFT operators contribute \cite{Azatov:2016sqh} 
at \dsix because the SM amplitude is CP-even and the integrated effect of interfering the SM amplitude with a CP-odd amplitude 
is zero.  This means that there is no contribution from the interference term to the inclusive rate, or to CP-even 
observables such as transverse momenta and invariant masses, and the only contribution is to appropriately constructed 
CP-odd observables.  This is not the case for terms proportional to $1/\Lambda^4$, which contain the squared \dsix 
amplitude and produce a CP-even effect regardless of the nature of the operator.  This has historically served as 
a motivation to constrain CP-odd operators with momentum-dependent observables in a range of production modes 
\cite{Plehn:2001nj,Hankele:2006ma, Englert:2012ct,Ellis:2012xd,Englert:2012xt,Dolan:2014upa, Casolino:2015cza,Goncalves:2018agy,Aad:2015tna,Aaboud:2018xdt,Aaboud:2017fye}. 
However, such an approach is more model-dependent since it neglects 
dimension-8 operators that interfere with the SM and in general produce similar ${\cal{O}}(1/\Lambda^4)$ effects. 

In this Letter we limit ourselves to interference-only effects so the constraints on CP-odd operators will be entirely 
derived from CP-odd observables, which are discussed in the next section. 
This approach is naturally less sensitive compared to including \dst~terms so it provides a conservative outlook into 
the future: if perturbatively meaningful constraints can be obtained in the linearised approach, these will only be 
strengthened if \dst~terms are included.

The interference-only contribution from each operator to each observable is constructed using {\sc{Madgraph5}}~\cite{Alwall:2014hca} 
and the SMEFT implementation of Ref.~\cite{Brivio:2017btx}. Event samples are produced separately for gluon-fusion and 
weak-boson-fusion production at fixed values of $c_i=1$ and $\Lambda=1$~TeV.  These parton-level events are passed to 
{\sc{Pythia8}}~\cite{Sjostrand:2014zea} to model the Higgs-boson decay, parton showering, hadronisation and multiple parton 
interactions. {\sc{Rivet}}~\cite{Buckley:2010ar} is then used to select events in each decay channel and to construct each 
observable according to the selection criteria published in the experimental papers. The cross-section contribution in each 
bin is multiplied by $\Gamma_{h\rightarrow XX} (c_i)/\Gamma_h(c_i)$, to account for the Higgs-boson branching fraction at 
the given point in EFT coupling space.  Interference-only predictions for each observable at other values 
of the Wilson coefficients are obtained by linear scaling.

All Standard Model predictions are taken from the experimental publications~\cite{Aaboud:2018xdt,Aaboud:2017oem}.  The gluon-fusion process was determined using {\sc{Powheg}} NNLOPS~\cite{Hamilton:2013fea} and scaled to the N$^3$LO inclusive cross section calculation with NLO electroweak corrections~\cite{Anastasiou:2015ema,Anastasiou:2016cez,Actis:2008ug,Anastasiou:2008tj}). For  vector boson fusion and Higgs boson production in association with a weak boson, the SM predictions were determined using 
{\sc{Powheg}}~\cite{Nason:2004rx,Alioli:2010xd,Nason:2009ai,Mimasu:2015nqa} and each were scaled to the NNLO calculation with NLO electroweak corrections applied~\cite{Ciccolini:2007jr,Ciccolini:2007ec,Bolzoni:2010xr,Brein:2003wg,Altenkamp:2012sx,Denner:2011id}.

\section{Framework and Fitting}
\label{sec:fit}

We implement our statistical tests by constructing a likelihood function $L(\boldsymbol{c}/\Lambda^2)$ for all the observables 
\begin{align}
 L(\boldsymbol{c}/\Lambda^2) = \prod_{i}^{\rm observables} L_{i}(\boldsymbol{c}/\Lambda^2) \, ,
 \label{eq:likelihood}
\end{align}
with $L_{i}(\boldsymbol{c}/\Lambda^2)$ denoting the likelihood of an individual observable $\boldsymbol{o}_{i}$ for a given vector 
of EFT coefficients $\boldsymbol{c}/\Lambda^2$.  We assume Gaussian uncertainties on the $h \to \gamma\gamma$ and $h \to 4 \ell$ 
differential cross-section measurements and express the likelihood as 
\begin{align}\label{eq:mult_var_gauss}
 L_i(\boldsymbol{c}/\Lambda^2) = 
\frac{1}{\sqrt{ 2\pi \sigma^2}} \exp\left( - \frac{\left( \boldsymbol{o}_{i} - \boldsymbol{\tau}_{i} \right)^2}{2\sigma^2} \right) \, ,
\end{align} 
with $\boldsymbol{\tau_{i}} = \boldsymbol{\tau_{i}}(\boldsymbol{c}/\Lambda^2)$ denoting the expected cross-section vector, which is 
constructed from the SM and interference-only cross-section contributions discussed in the previous section.  Estimators 
$(\hat{\boldsymbol{c}}/\Lambda^2)$ for the Wilson coefficients are obtained by numerically maximising $L$ to obtain $L_{\rm max}$, and 
confidence intervals (CI) are constructed using the asymptotic behaviour of the likelihood.  The CI are defined by finding 
value(s) of $\boldsymbol{c}/\Lambda^2$ such that for a fixed CI
\begin{align}
 1 - \text{CI} = \int_{- 2 \ln L( \boldsymbol{c}/\Lambda^2 ) + 2 \ln L_{\rm max}}^{\infty} \,  f_{\chi^2}(x; m \, \text{dof}) \, \text{d} x \, ,
\end{align}
with $f_{\chi^2}(x; m \, \text{dof})$ denoting the $\chi^2$-distribution with $m = \text{dim}( \boldsymbol{c} )$ degrees 
of freedom.   

The likelihood function is implemented in the \textsc{GammaCombo} package~\cite{Aaij:2016kjh}, which uses \textsc{Minuit} to carry 
out the numerical maximisation and relevant profiling. The two-dimensional coverage of the shown results correspond to 68.3\% and 
95.5\% CI.  The level of bias in the estimators $\boldsymbol{c}/\Lambda^2$ and the accuracy of the coverage have been tested using 
ensembles of pseudo-experiments generated around the SM and benchmark points.

\section{Results with existing measurements}
\label{sec:results_curr}

The most constraining model-independent Higgs boson measurements are the differential cross sections in the $h\to \gamma\gamma$ 
and $h\to ZZ^* \rightarrow 4\ell$ decay channels. In this analysis we use recent ATLAS measurements made at $\sqrt{s}=13$~TeV~\cite{Aaboud:2018xdt,Aaboud:2017oem}. The differential 
cross sections published by CMS~\cite{Sirunyan:2018kta,Sirunyan:2017exp}, and by ATLAS in the $h\to WW^* \to \ell\nu\ell\nu$ decay channel~\cite{Aad:2016lvc}, do not include observables sensitive to 
CP-odd interference effects and are therefore not included in our combination. 
As yet, differential cross sections have not been published for any other Higgs boson decay channels. 

Of the distributions measured in the $h\to \gamma\gamma$ and $h\to ZZ^* \rightarrow 4\ell$ decay channels, only the signed $\Delta \phi_{jj}$ 
between the two jets in $h+2~{\rm jet}$ events is a CP-sensitive observable. The signed $\Delta \phi_{jj}$ probes the CP structure of the Higgs boson's interaction with gluons or weak bosons in the gluon-fusion~\cite{Hankele:2006ma,Hankele:2006ja,Klamke:2007cu} (see also \cite{Figy:2004pt}) and vector-boson fusion~\cite{Plehn:2001nj} production mechanisms, respectively, and is defined as 
\begin{equation}
\Delta \phi_{jj} = \phi_1 - \phi_2,
\end{equation}
where $\phi_{1,2}$ are the azimuthal angles of the two jets with the highest transverse momentum ($p_{\rm T}$) in the event, ordered according to their rapidities ($y$) such that $y_1>y_2$.  The 
asymmetry in the signed-$\Delta\phi_{jj}$ distribution is a model-independent probe of CP-violation.\footnote{Asymmetries in $\Delta \phi_{jj}$ can result either from a CP-odd operator or from an absorptive phase of a CP-even operator~\cite{Brehmer:2017lrt}.  We take all Wilson coefficients to be real and extract constraints on the CP-odd operators; the constraints can be more generally interpreted as applying to the combination of the CP-odd operators and the imaginary coefficients of the CP-even operators.} and is defined as
\begin{equation*}
A = \frac{\sigma(0<\Delta\phi_{jj}<\pi) - \sigma(-\pi<\Delta\phi_{jj}<0)}{\sigma(0<\Delta\phi_{jj}<\pi) + \sigma(-\pi<\Delta\phi_{jj}<0)},
\end{equation*}
where $\sigma$ is the measured fiducial cross section in each region of $\Delta\phi_{jj}$. The asymmetry obtained by statistically combining the ATLAS data in the $h\to \gamma\gamma$ and $h\to ZZ^* \rightarrow 4\ell$ decay channels is $0.3 \pm 0.2$.\footnote{The measurement have uncertainties that are dominantly statistical, so correlations in systematic uncertainties are expected to have a negligible impact on the combined asymmetry.  The signed-$\Delta\phi_{jj}$ distribution in the $h\to ZZ^* \rightarrow 4\ell$ measurement was originally presented in the range $0<\Delta\phi_{jj}<2\pi$ and is transformed accordingly for the analysis presented here.}  If the central value were to persist in future high-precision measurements made with larger datasets, it could be an indication of non-SM CP-violation in the Higgs sector. 

The global analysis framework discussed in Sec.~\ref{sec:fit} is used to characterise the possible source of the modest asymmetry.  All 
four CP-odd operators in Eq.~\eqref{eq:ops} can produce an asymmetry in the signed $\Delta \phi_{jj}$ distribution.  Table~\ref{tab:constr} shows the one-dimensional 95\% confidence level constraints from a fit to the measured ATLAS differential $\Delta\phi_{jj}$ distributions.
\begin{table}[!tbp]
\centering
\begin{tabular}{|c|c|}
\hline\hline
Coefficient $\left[\mathrm{TeV}^{-2}\right]$ & Constraint \\
\hline
\rule{0pt}{\normalbaselineskip} 
$c_{H\tilde{G}}/\Lambda^2$ & [ -0.19 , 0.03 ] \\ 
$c_{H\tilde{W}}/\Lambda^2$ & [ -3 , 16 ] \\ 
$c_{H\tilde{B}}/\Lambda^2$ & [ -2900 , 3300 ] \\ 
$c_{H\tilde{W}B}/\Lambda^2$ &  [ -730 , 160 ] \\[0.5ex] 
\hline \hline
\end{tabular}
\caption{\label{tab:constr} The 95\% confidence interval for each Wilson coefficient, in units of TeV$^{-2}$, when all other Wilson coefficients are set to zero.
}
\end{table}

The signed $\Delta \phi_{jj}$ distribution is mainly sensitive to the $O_{H\tilde{G}}$ and $O_{H\tilde{W}}$ operators, with little 
sensitivity to the other CP-odd operators. This is because only the $O_{H\tilde{G}}$ operator affects gluon-fusion production, and $WWh$ interactions dominate the weak-boson-fusion contribution.  To constrain the $O_{H\tilde{B}}$ and $O_{H\tilde{W}B}$ operators we need observables dominated by Higgs boson interactions with the $Z$ boson; we will discuss such observables in Sec.~\ref{sec:results_proj}.

\begin{figure}[!t]
\includegraphics[width=0.46\textwidth]{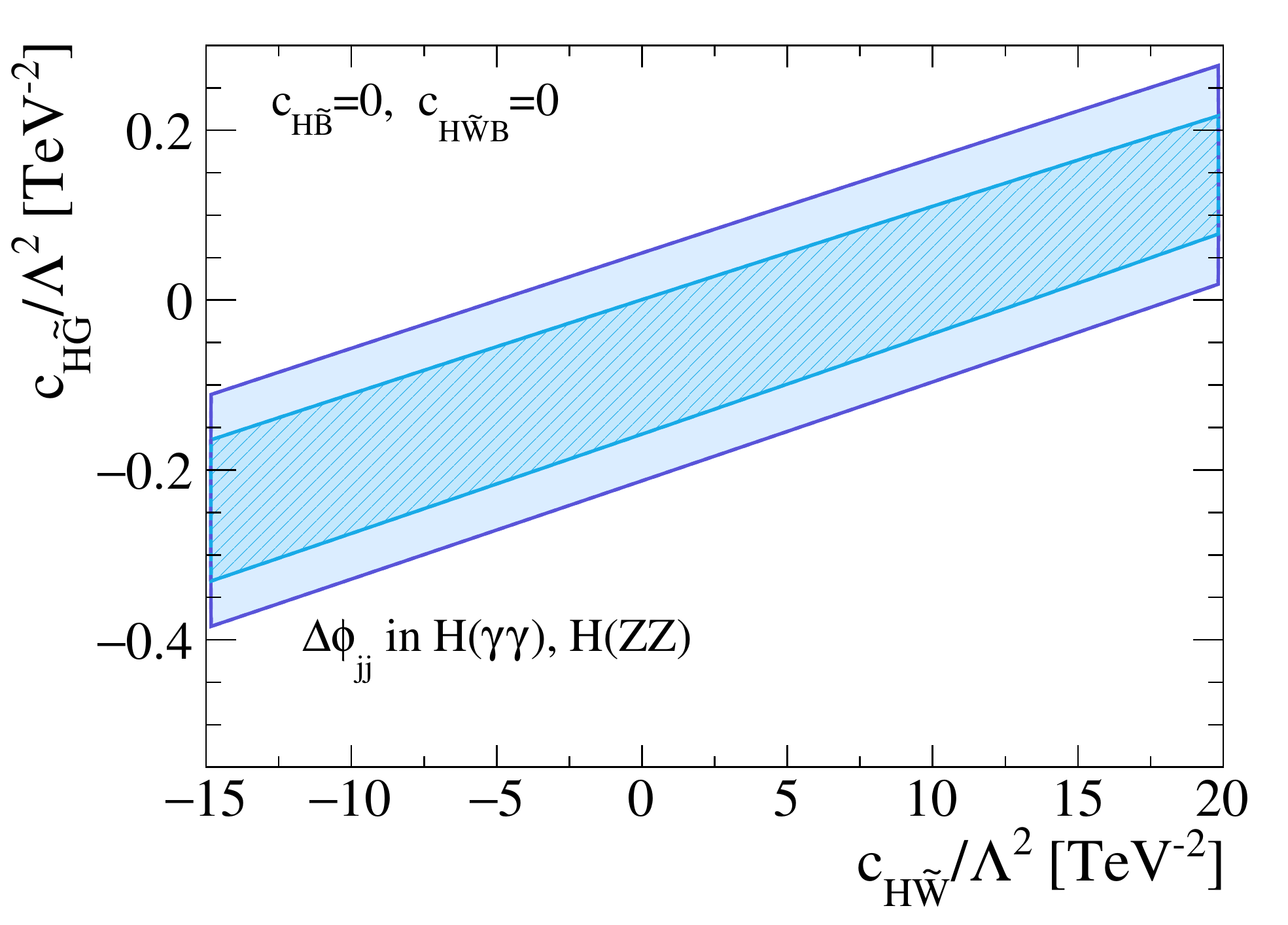}
\caption{\label{fig:phijjodd} Constraints on the coefficients of two CP-odd operators from data when setting all other Wilson coefficients to zero. Contours are presented when using only the signed $\Delta\phi_{jj}$ distribution. 
Inner and outer shaded regions represent the 68.3\% and 95.5\% CI, respectively.
}
\end{figure}

The correlation between the leading operators affecting gluon fusion and vector-boson fusion are shown in Fig.~\ref{fig:phijjodd}.  The constraints are highly correlated since the operators can not be distinguished just from the asymmetry of the $\Delta \phi_{jj}$ distribution.  ATLAS measures the distribution in four bins in the $h\to \gamma\gamma$ channel so it provides a modest discrimination of the $c_{H\tilde{G}}$ and $c_{H\tilde{W}}$ operators. 
The correlation between these operators could be further reduced by separating the $\Delta \phi_{jj}$ measurements into regions that enhance either gluon fusion or vector-boson fusion.  We will discuss such a possibility in Sec.~\ref{sec:results_proj}.  

\begin{figure}[th]
\includegraphics[width=0.46\textwidth]{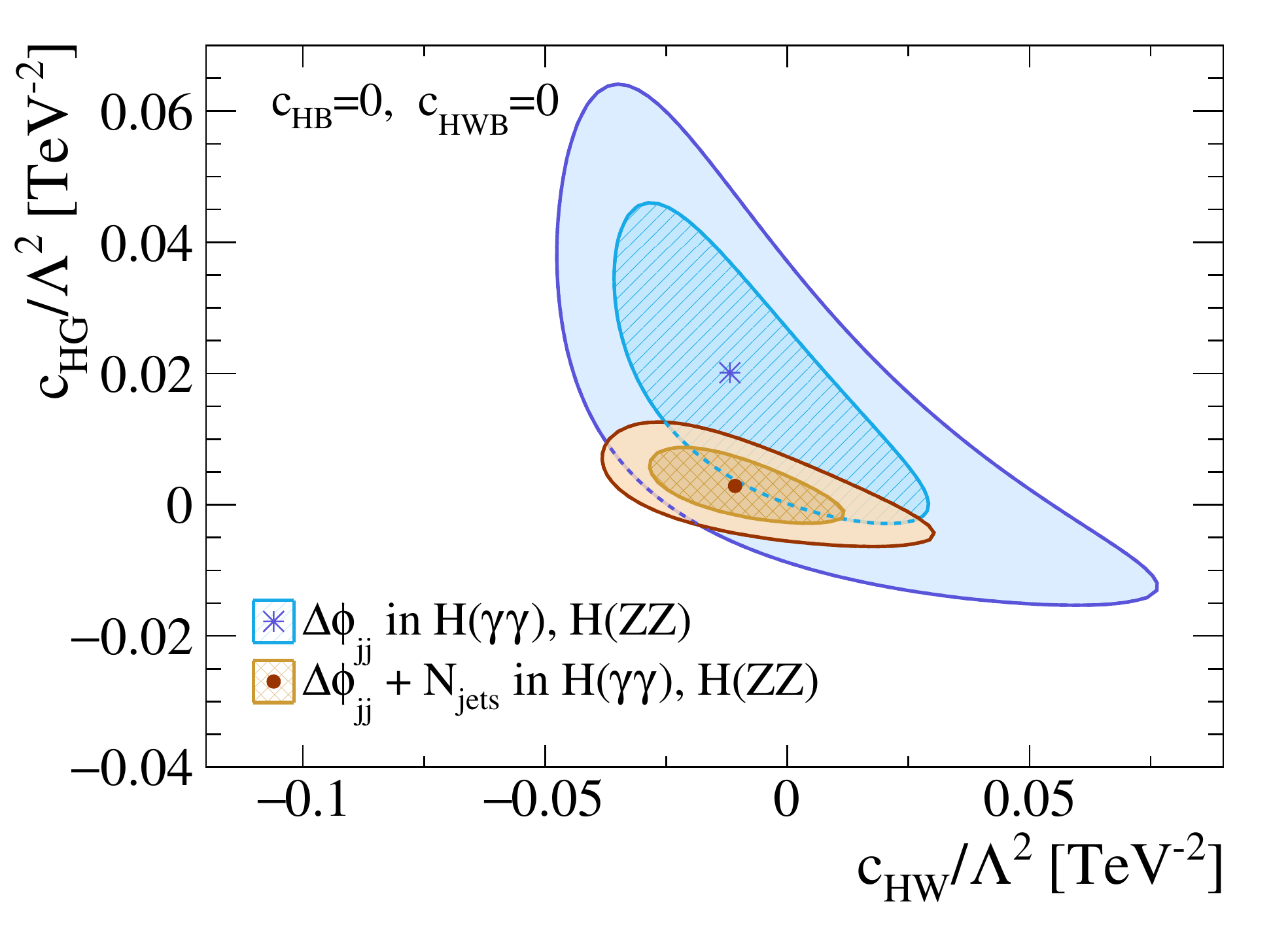}
\hspace{1cm}
\includegraphics[width=0.46\textwidth]{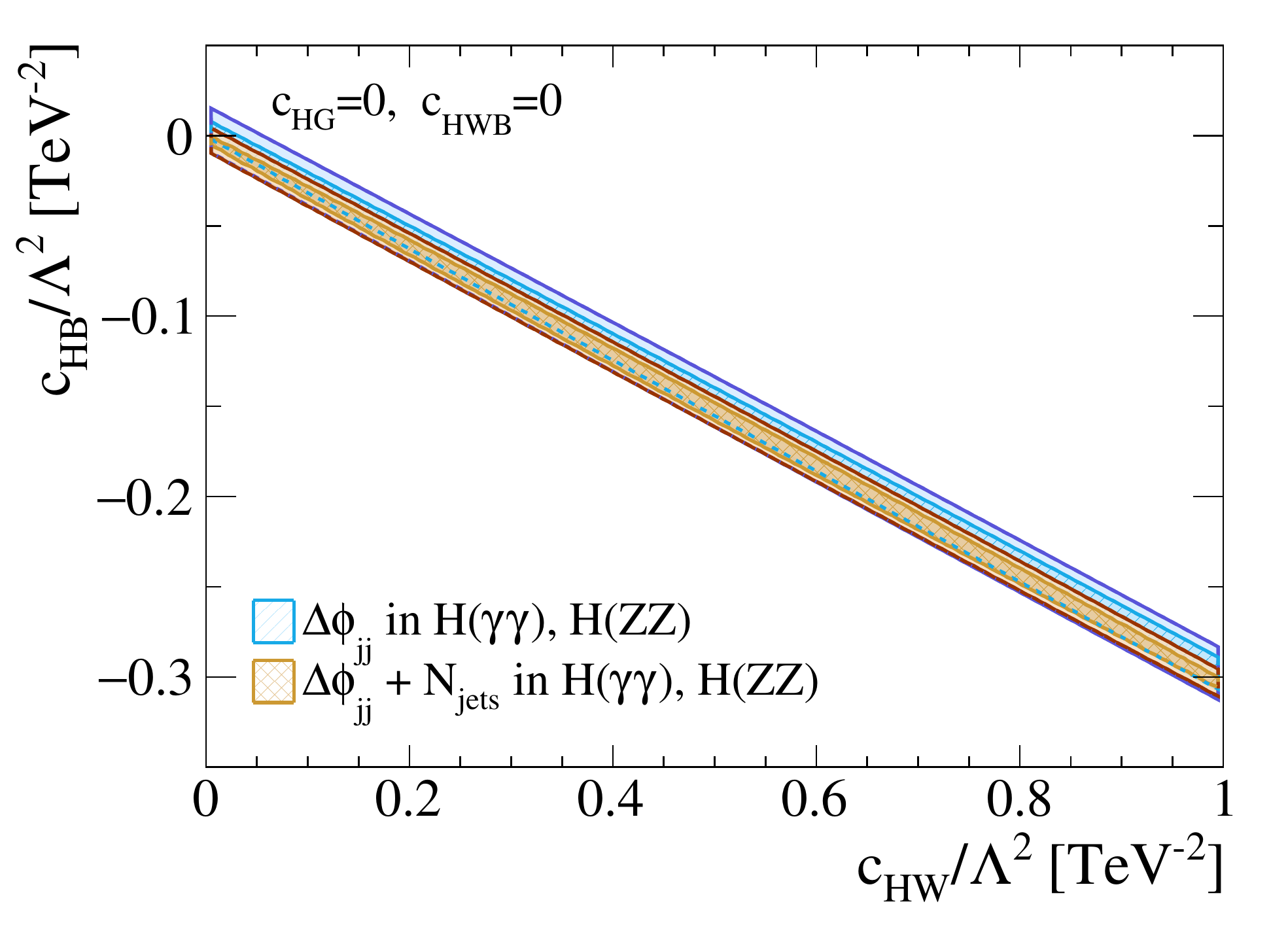}
\caption{\label{fig:phijjeven} Constraints on the CP-even operators from $\Delta\phi_{jj}$ measurements only, and from a combination 
of the $\Delta\phi_{jj}$ and jet-multiplicity measurements. The best-fit points are shown within the contours. 
Inner and outer shaded regions represent the 68.3\% and 95.5\% CI, respectively.}
\end{figure}

Finally, although CP-even observables do not help with understanding possible sources of CP-violation, they provide information 
on associated operators in the EFT.  If one of the CP-odd operators in Table~\ref{tab:constr} were shown to be non-zero, one would 
clearly want to know the allowed values of the corresponding CP-even operator.  Figure~\ref{fig:phijjeven} shows the constraints 
on the CP-even operators that can be obtained from the $\Delta \phi_{jj}$ distribution, and the significant additional constraining 
power of the jet multiplicity distribution ($N_\mathrm{jets}=0,1$) measured by ATLAS in the $h\to \gamma\gamma$ and $h\to ZZ^* \rightarrow 4\ell$ decay channels.  There is good agreement between the data and predictions for both the SM and EFT hypotheses in all the fits, as demonstrated by the
$\chi^2$ values shown in Table~\ref{tab:chi2}.

\begin{table}[!htbp]
\centering
\begin{tabular}{|c|c|c|}
\hline\hline
Coefficient(s) & Variable(s) & $\chi^2$ / ndf \\
\hline
\rule{0pt}{\normalbaselineskip} 
$c_{H\tilde{G}}/\Lambda^2$ &$\Delta \phi_{jj}$  & 3.5 / 5 \\
$c_{H\tilde{W}}/\Lambda^2$ &$\Delta \phi_{jj}$ & 3.6 / 5 \\
$c_{H\tilde{B}}/\Lambda^2$ & $\Delta \phi_{jj}$& 5.5 / 5 \\
$c_{H\tilde{W}B}/\Lambda^2$ &$\Delta \phi_{jj}$ & 3.9 / 5  \\
$c_{H\tilde{W}}/\Lambda^2:c_{H\tilde{G}}/\Lambda^2$ & $\Delta \phi_{jj}$&  3.3 / 4  \\
$c_{H\tilde{W}B}/\Lambda^2:c_{H\tilde{W}}/\Lambda^2$ &$\Delta \phi_{jj}$ &  3.5 / 4 \\
$c_{HW}/\Lambda^2:c_{HG}/\Lambda^2$ & $\Delta \phi_{jj}$, $N_{\rm jets}$ &  7.6 / 8  \\
$c_{HW}/\Lambda^2:c_{HB}/\Lambda^2$ &$\Delta \phi_{jj}$, $N_{\rm jets}$ &  6.5 / 8 \\[0.5ex] 
\hline \hline
\end{tabular}
\caption{\label{tab:chi2} Summary of the fit quality for all one-dimensional and two-dimensional fits of Wilson coefficients to existing measurements. The $\chi^2$ values are calculated from the maximised likelihood shown in Eq.~\ref{eq:likelihood}. The corresponding SM hypotheses have  $\chi^2_{\rm SM}/\textrm{ndf} = 5.5/6$ and  $\chi^2_{\rm SM}/\textrm{ndf} = 8.2/10$ for fits to the $\Delta \phi_{jj}$ data only and to the combined $\Delta \phi_{jj}$ and jet multiplicity ($N_\mathrm{jets}$) data, respectively.}
\end{table}

In contrast to the CP-odd operators, the net effect of the interference between ${\cal{M}}_\text{SM}$ and ${\cal{M}}_\text{d6}$ is non-zero for CP-even operators, so rate and kinematic information can be used to significantly constrain the corresponding Wilson coefficients. Figure~\ref{fig:phijjeven} (top) shows that the $O_{HG}$ operator can be untangled easily from $O_{HW}$, because operators affecting gluon fusion production have the same impact in both decay channels, whereas operators affecting the Higgs interaction with weak bosons are most tightly constrained by the $h\rightarrow \gamma \gamma$ branching ratio. Blind directions still exist, however, when constraining two operators that affect the $h\rightarrow \gamma \gamma$ branching ratio, as shown in Fig.~\ref{fig:phijjeven} (bottom).  
In a global analysis with other Higgs-boson measurements, all of these operators will be more tightly constrained~\cite{Ellis:2018gqa,Brehmer:2016nyr}. We emphasise that the ability to constrain the CP-odd couplings using CP-sensitive observables will not be affected by blind directions in the CP-even coupling space.

\section{Enhancing the sensitivity to CP-violation in the Higgs sector}
\label{sec:results_proj}

The results of the fit to existing data raise the question of how we can improve sensitivity to CP-odd effects through targeted 
measurements. In particular, the current ATLAS $\Delta\phi_{jj}$ measurements do not distinguish between CP-violating interactions in gluon fusion and vector-boson fusion production of the $h+2~{\rm jet}$ final state. This degeneracy can be trivially removed by separating the measurement into regions that enhance either gluon fusion or vector-boson fusion. 
ATLAS have constrained CP-odd operators that impact vector-boson fusion in a VBF-enhanced phase space in the $h\to \tau\tau$ decay channel~\cite{Aad:2016nal}. However, CP-odd operators that impact gluon fusion were not considered and the CP-sensitive observables were not presented in a well-defined fiducial region. We are therefore not able to include the results in our combination.

It is also important to address the lack of sensitivity to the $O_{H\tilde{B}}$ and $O_{H\tilde{W}B}$ operators. These operators 
can be probed through the study of angular production and decay observables in Higgs boson production
processes \cite{DellAquila:1985mtb,DellAquila:1985jin,Trueman:1978kh,Cabibbo:1965zzb,Collins:1977iv,Nelson:1986ki,Miller:2001bi,Choi:2002jk,DeRujula:2010ys,Gao:2010qx,Choi:2012yg,Bolognesi:2012mm}. 
For the $h \rightarrow ZZ^\ast\rightarrow \ell^+ \ell^- \ell^{\prime +}\ell^{\prime -}$ system, an angle that is particularly sensitive to CP is the 
$\Phi$ variable \cite{Bolognesi:2012mm} defined through
\begin{equation}
\label{eq:decayplane}
\cos\Phi = { ( {\vec{p}}_{\ell^-} \times {\vec{p}}_{\ell^+} ) \cdot ( {\vec{p}}_{\ell^{\prime -}} \times {\vec{p}}_{\ell^{\prime +}} ) \over
\sqrt{( {\vec{p}}_{\ell^-}  \times {\vec{p}}_{\ell^+}  )^2\,  ( {\vec{p}}_{\ell^{\prime -}} \times {\vec{p}}_{\ell^{\prime +}} )^2 }}  \bigg|_h\,,
\end{equation}
calculated in the Higgs boson centre-of-mass frame. This observable
could already be measured with existing data. Decay angles have been used by both ATLAS and CMS to search for CP-violation in the $h\rightarrow ZZ^\ast\rightarrow 4\ell$ and $h\rightarrow WW^\ast\rightarrow \ell\nu\ell\nu$ decay channels~\cite{Aad:2015mxa,Sirunyan:2017tqd}. However, in 
these searches, the detector-level data were analysed using either boosted decision trees or matrix-element-based likelihood analyses and 
the results cannot be interpreted in terms of the CP-odd operators we consider.  The results are consistent with zero CP-asymmetry. 

The impact that additional measurements could have in a global analysis is studied using pseudo-data assuming 36/fb of integrated 
luminosity at $\sqrt{s}=13$~TeV. In both the $h\to ZZ^* \to 4\ell$ and $h\to \gamma\gamma$ decay channels, the pseudo-data are constructed for the signed $\Delta\phi_{jj}$ using the SM expectation and the measured uncertainties in data, since the measurements are dominated by either signal or background statistical uncertainties.  A two-bin signed $\Delta\phi_{jj}$ distribution is constructed in VBF-enhanced and VBF-suppressed regions in the $h\to \gamma\gamma$ channel, using the published differential cross sections and SM expectations for the $N_{\rm jet}\geq2$ and {\it VBF-enhanced} phase spaces from~\cite{Aaboud:2018xdt}.   

\begin{figure*}[!htbp]
\subfigure[ Constraints on $O_{H\tilde{G}}$ and $O_{H\tilde{W}}$]{
\includegraphics[width=0.46\textwidth]{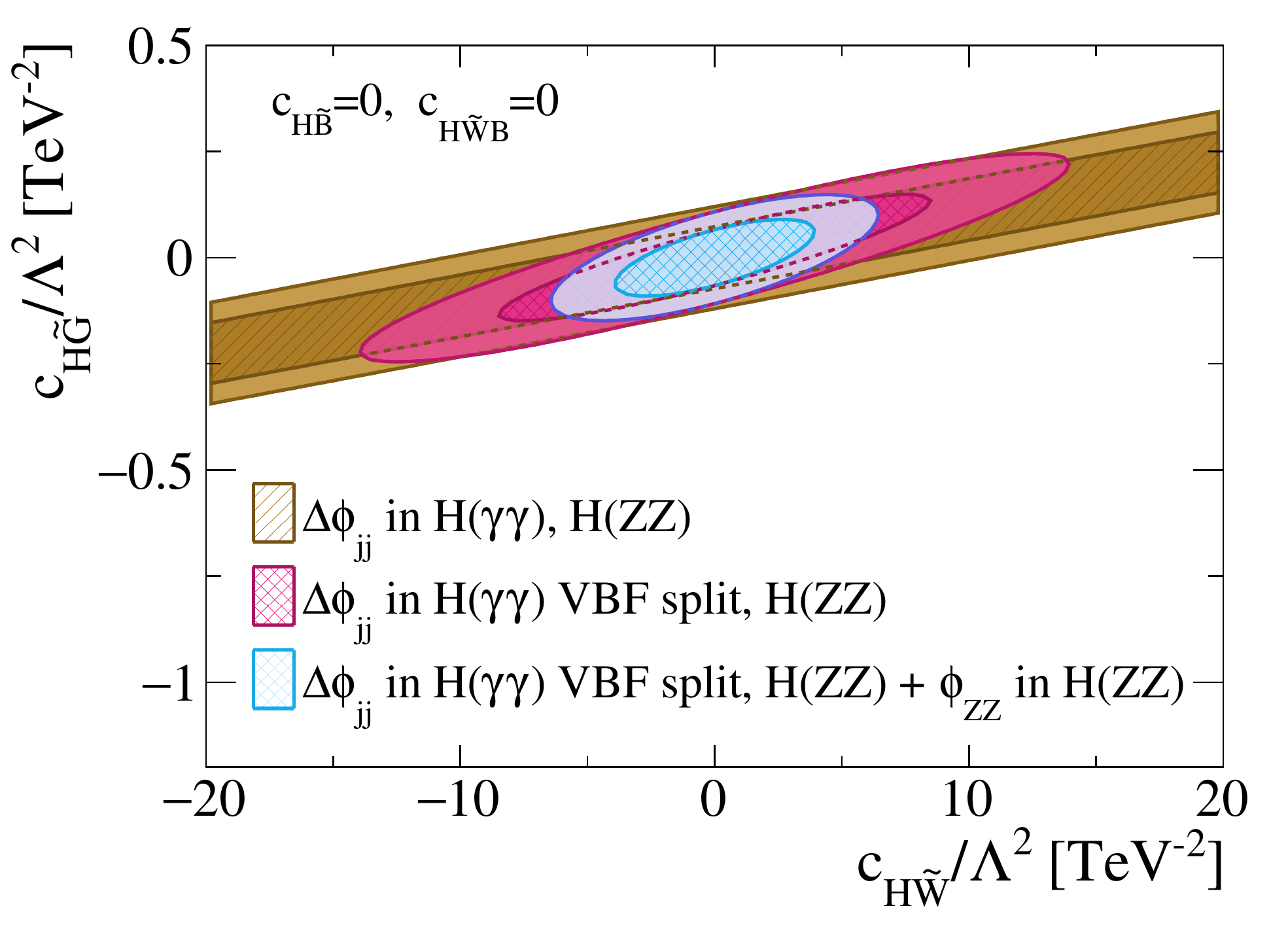}}
\hspace{1cm}
\subfigure[ Constraints on $O_{H\tilde{W}}$ and $O_{H\tilde{B}}$]{
\includegraphics[width=0.46\textwidth]{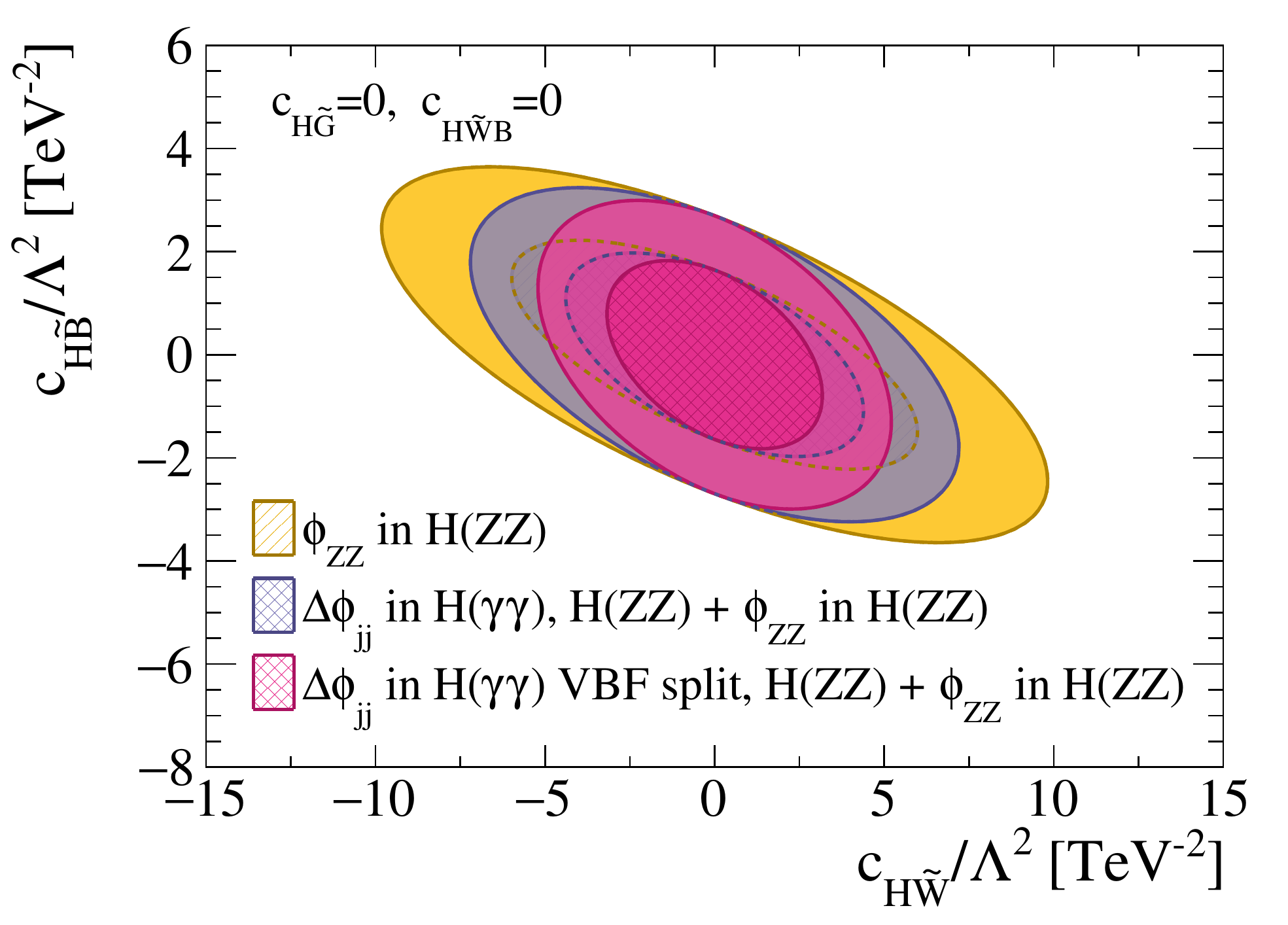}}
\caption{\label{fig:phijjoddps}  (a) Individual constraints on the coefficients of the leading CP-violating operators affecting 
gluon fusion ($O_{H\tilde{G}}$) and vector-boson fusion ($O_{H\tilde{W}}$). The blind direction resulting from inclusive 
$\Delta\phi_{jj}$ measurements is resolved through the use of VBF-enhanced and VBF-suppressed kinematic regions.  (b) Individual constraints on two CP-violating interactions affecting vector-boson fusion 
($O_{H\tilde{W}}$ and $O_{H\tilde{B}}$).
The results are obtained using pseudo-data, and the inner and outer shaded regions represent the 68.3\% and 95.5\% CI, respectively.}
\end{figure*}

The results of the global analysis of the pseudo-data are shown in Fig.~\ref{fig:phijjoddps}(a) when constraining the $O_{H\tilde{G}}$ 
and $O_{H\tilde{W}}$ operator coefficients, with all other Wilson coefficients set to zero. It is clear that these operators can be distinguished by appropriate measurements of signed-$\Delta\phi_{jj}$ in VBF-enhanced
 and VBF-suppressed phase spaces, and the constraints are further improved with the addition of the $\Phi$ decay-angle observable in events with $N_{\rm jet} < 2$. 
Furthermore, the addition of the $\Phi$ variable allows the extraction of the $O_{H\tilde{B}}$ or $O_{H\tilde{W}B}$ coefficient.  
Figure~\ref{fig:phijjoddps}(b) shows the constraints on the $O_{H\tilde{B}}$ coefficients using the decay angle information 
alone, and the improvement in the 2D plane when the signed $\Delta\phi_{jj}$ information is added.

\begin{figure*}[!htbp]
\subfigure[]{
\includegraphics[width=0.46\textwidth]{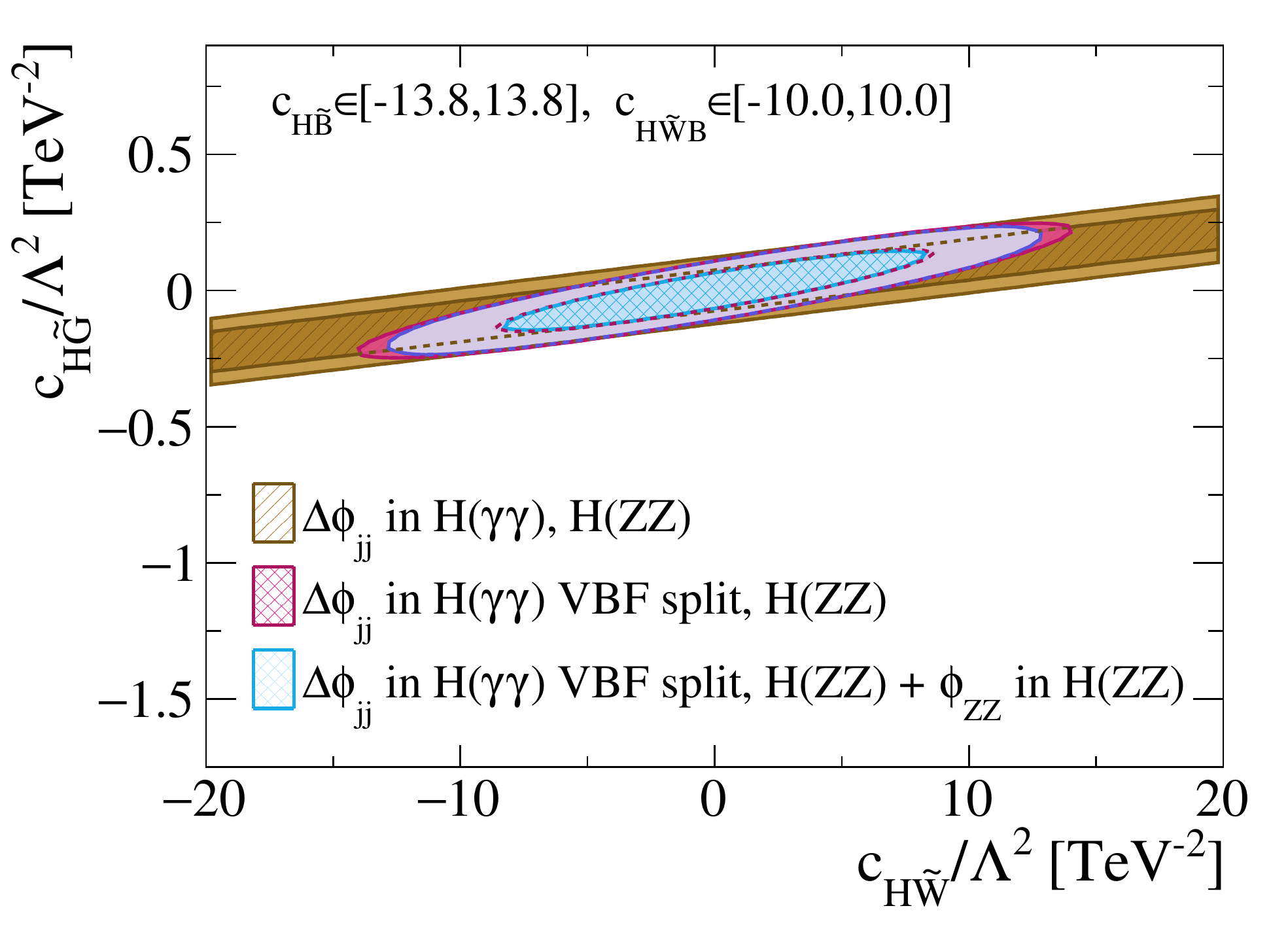}}
\hspace{1cm}
\subfigure[]{
\includegraphics[width=0.46\textwidth]{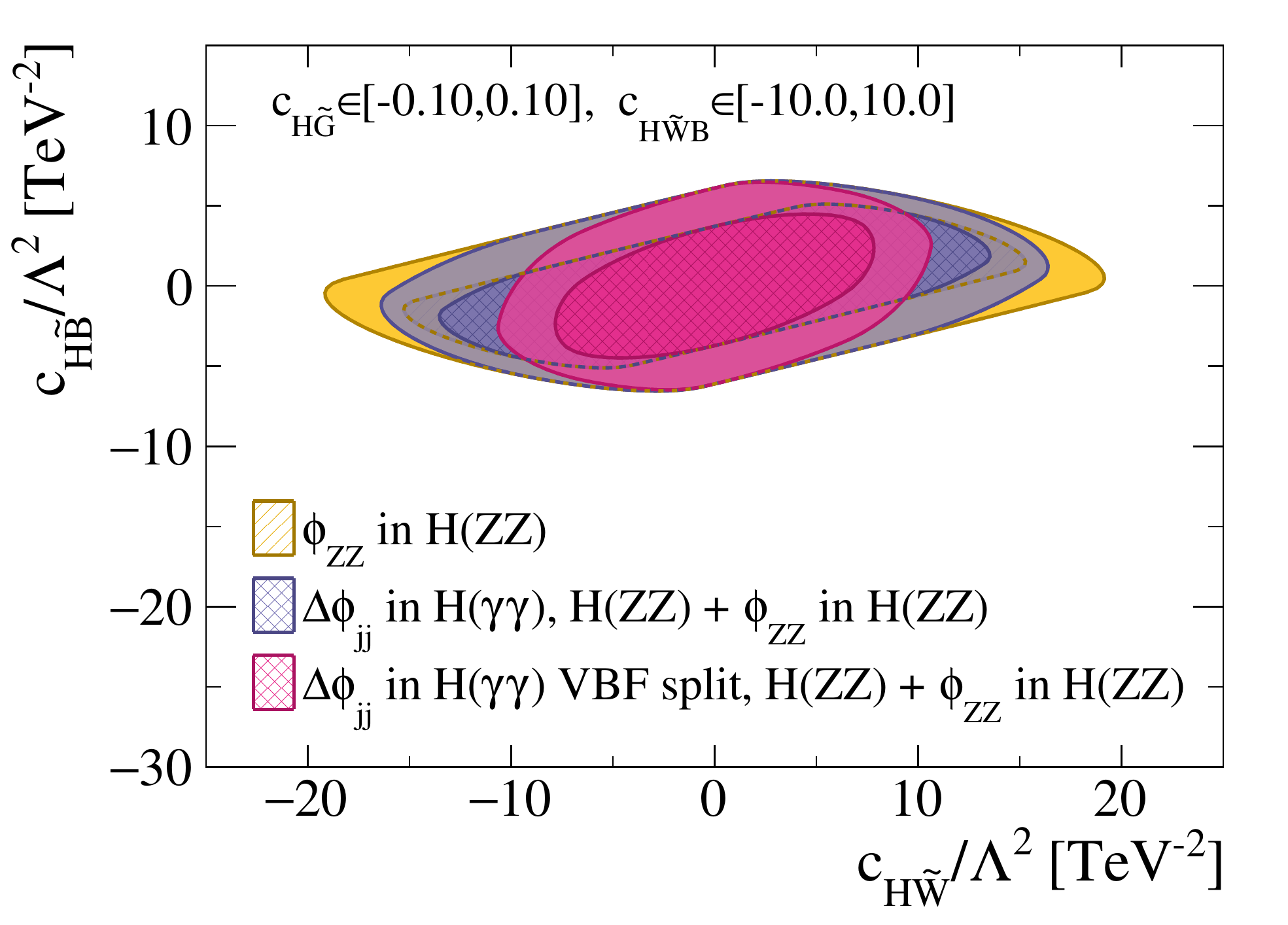}}
\subfigure[]{
\includegraphics[width=0.46\textwidth]{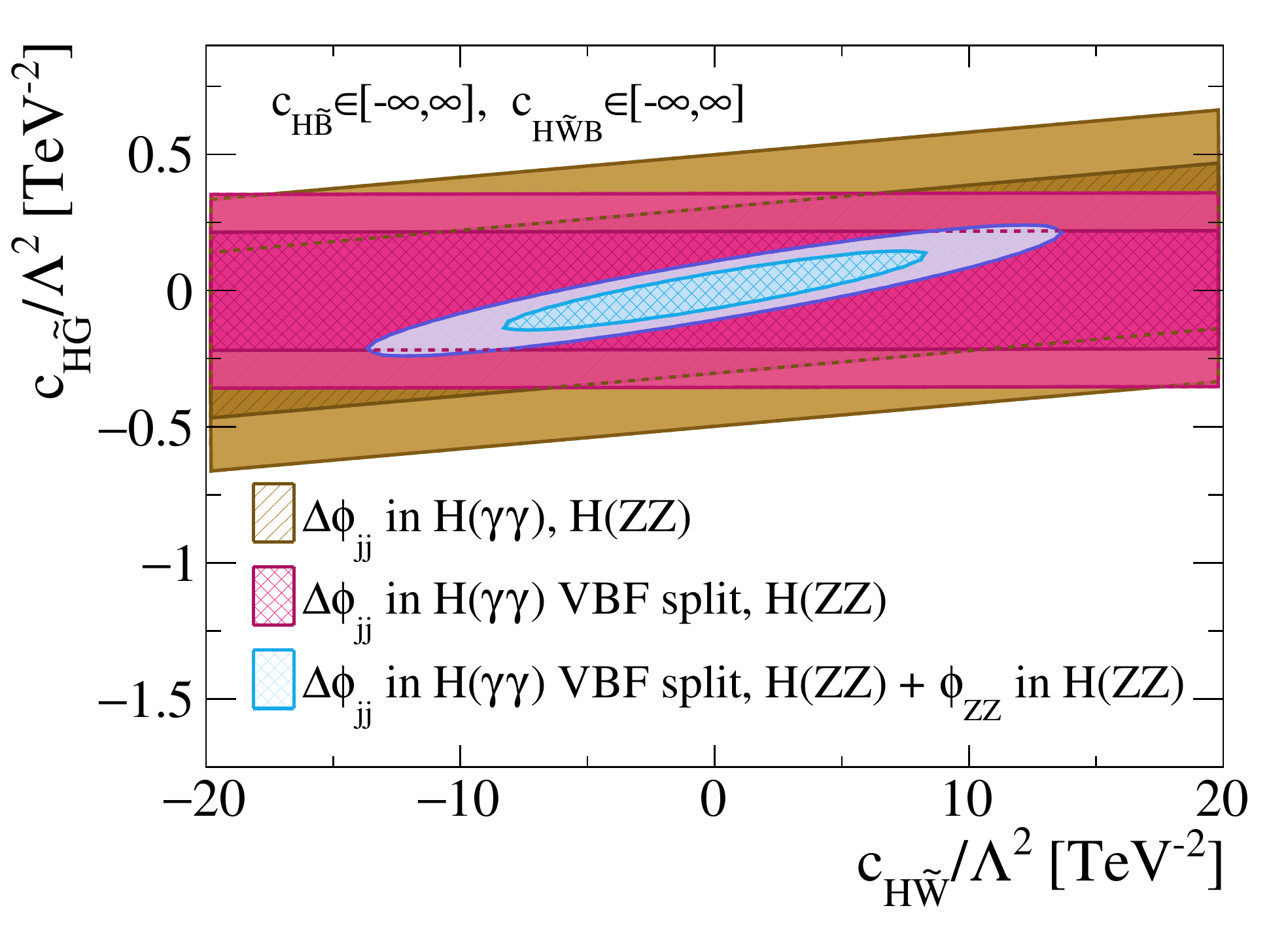}}
\hspace{1cm}
\subfigure[]{
\includegraphics[width=0.46\textwidth]{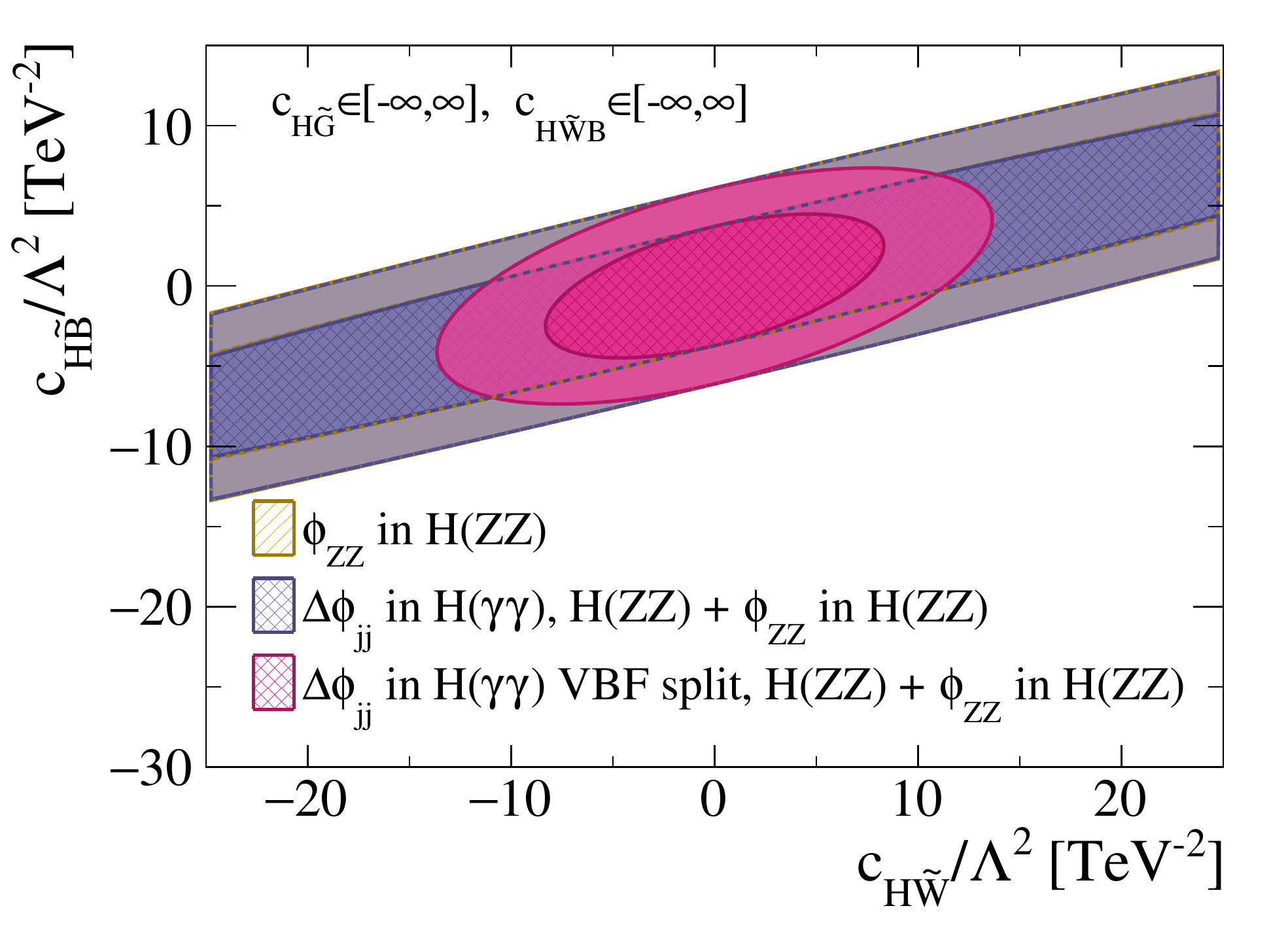}}
\caption{\label{fig:phijjoddps_marg} Individual constraints on two CP-violating interactions 
affecting gluon fusion ($O_{H\tilde{G}}$) and vector-boson fusion ($O_{H\tilde{W}}$) (panels (a) and (c)) and affecting
vector-boson fusion ($O_{H\tilde{W}}$ and $O_{H\tilde{B}}$) (panels (b) and (d)). 
Panels (a) and (b) show the 2D constraints after marginalising over other CP-odd operators with the constraint that the
associated Wilson coefficients satisfy the condition in Eq.~\eqref{eqn:eft_valid}.  Panels (c) and (d)
show the same 2D constraints after marginalisation over other CP-odd operators, with no conditions on the size of
the Wilson coefficients. 
The results are obtained using pseudo-data, and inner and outer shaded regions represent the 68.3\% and 95.5\% CI, respectively.}
\end{figure*}

The combination of all CP-sensitive observables is important when constraining all operators 
simultaneously.  To demonstrate this we recalculate the 2D constraints after marginalising over the other CP-odd operators. 
The marginalisation is subject to a perturbativity constraint such that 
\begin{equation}
\label{eqn:eft_valid}
\sum_i |\sigma_{\textrm{BSM}\times\textrm{SM}}^i| / \sigma_{\textrm{SM}} < 0.5,
\end{equation}
where $\sigma^i_{\textrm{BSM}\times\textrm{SM}}$ is the cross section of the interference term in 
bin $i$ of the observable.\footnote{The modulus is taken to avoid cancellation that would otherwise result from
summing across all bins of the measured observable. }  
This requirement ensures
that potential $(c_i/\Lambda^2)^2$ contributions to the interference term
$2\text{Re}\left( {\cal{M}}_\text{SM}^\star {\cal{M}}_\text{d6}\right)$, which include diagrams with two dimension-6 vertices, will be smaller than the 
leading term $c_i/\Lambda^2$ that we consider in our analysis.
With the current data the marginalisation over parameters within the perturbativity constraint 
does not have a significant effect, as shown in the top plots of Fig.~\ref{fig:phijjoddps_marg}.  If we drop this constraint 
the blind directions are clear (bottom row of Fig.~\ref{fig:phijjoddps_marg}), showing that as the measurements improve the 
combination of observables will become more important.

\begin{figure*}[!thbp]\subfigure[]{
\includegraphics[width=0.46\textwidth]{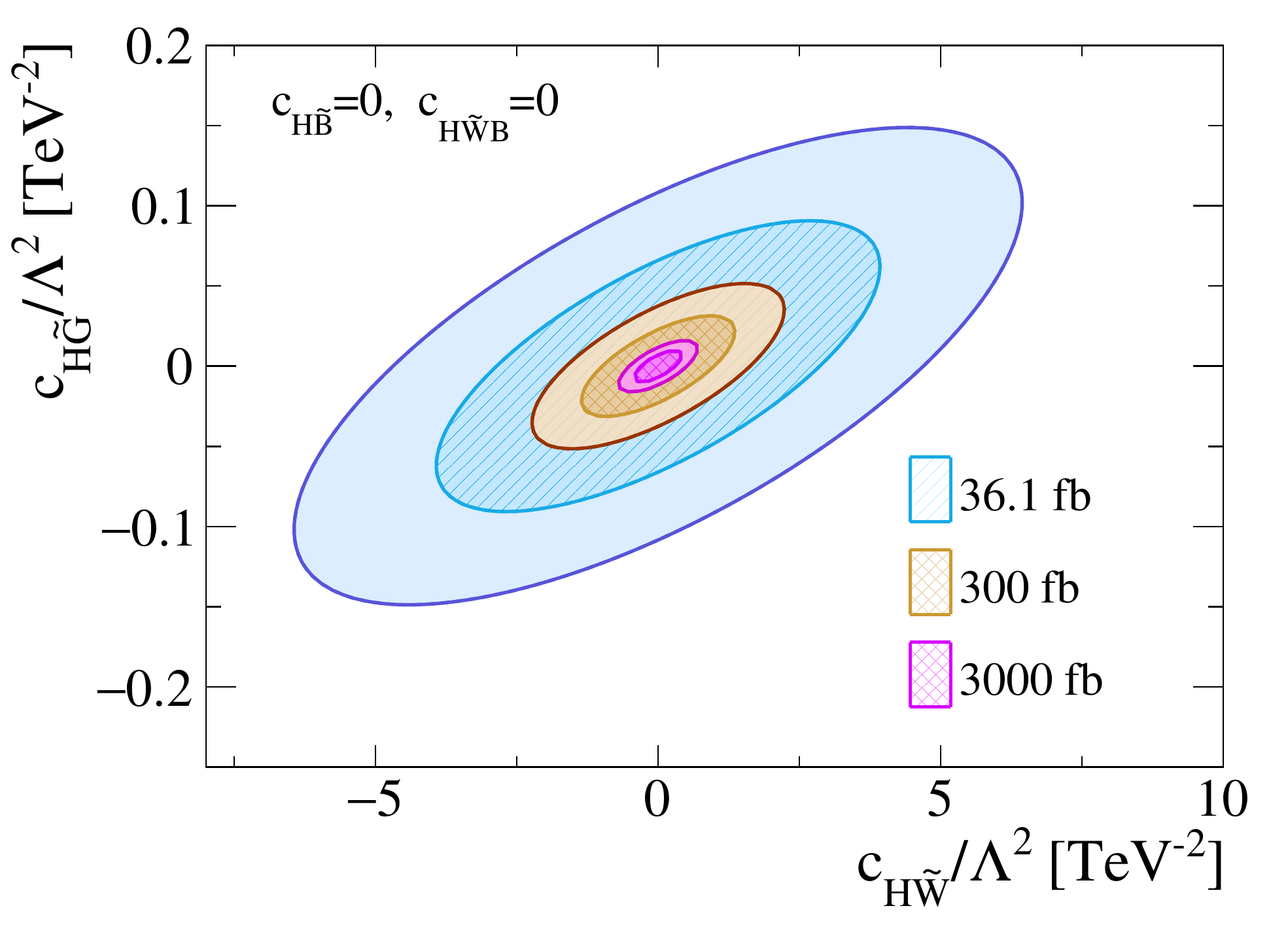}}
\hspace{1cm}\subfigure[]{
\includegraphics[width=0.46\textwidth]{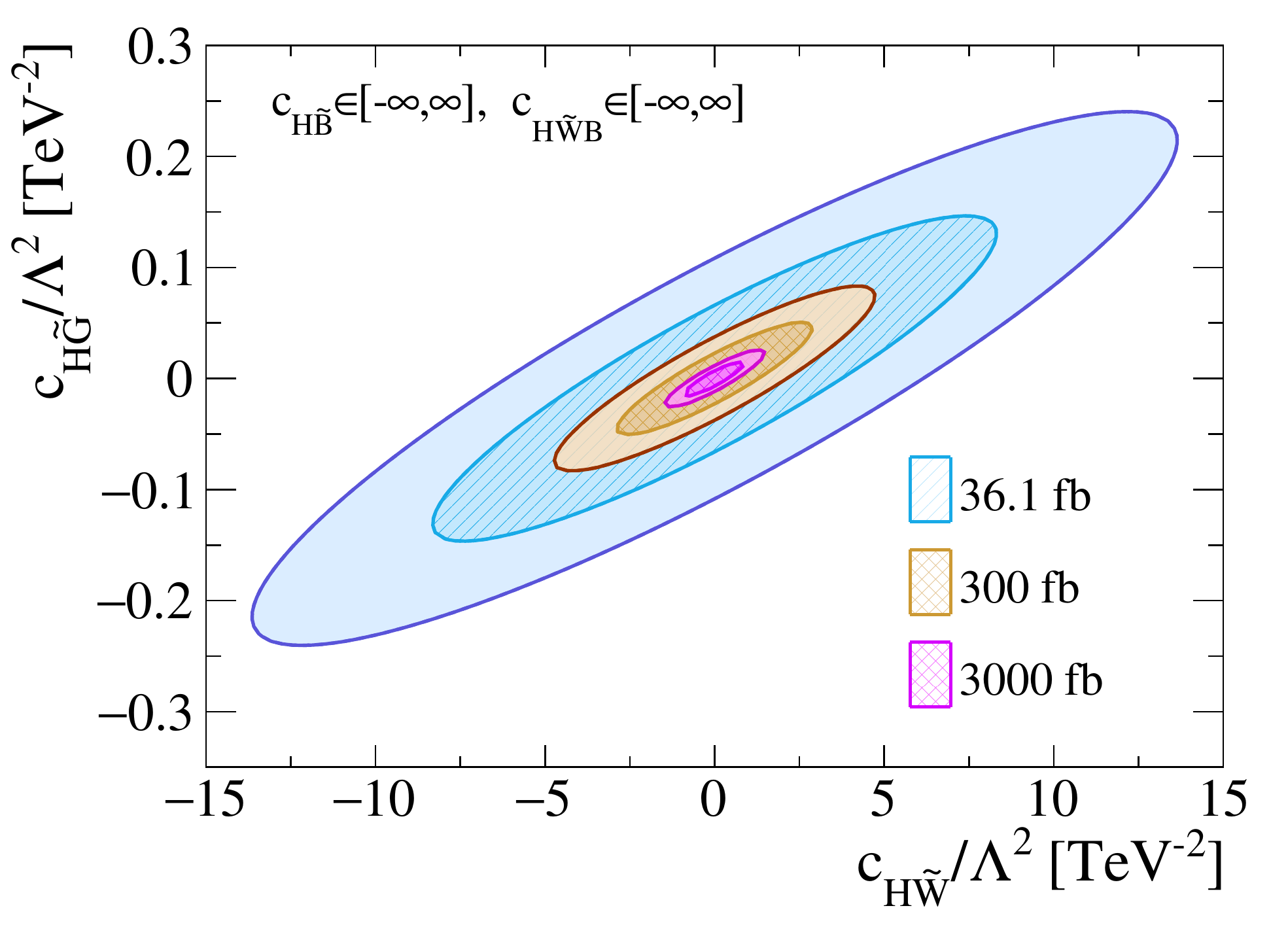}}
\subfigure[]{
\includegraphics[width=0.46\textwidth]{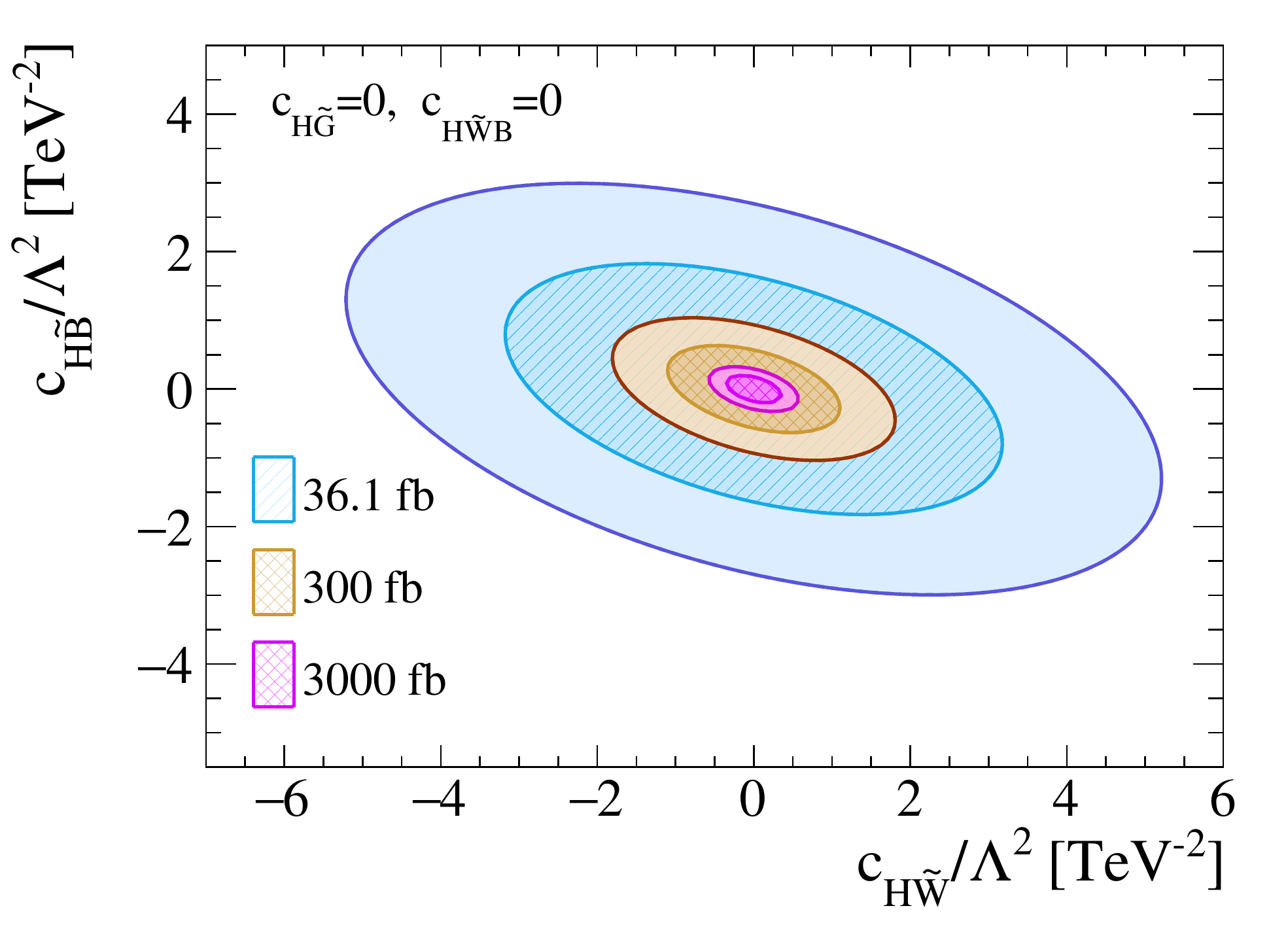}}
\hspace{1cm}
\subfigure[]{
\includegraphics[width=0.46\textwidth]{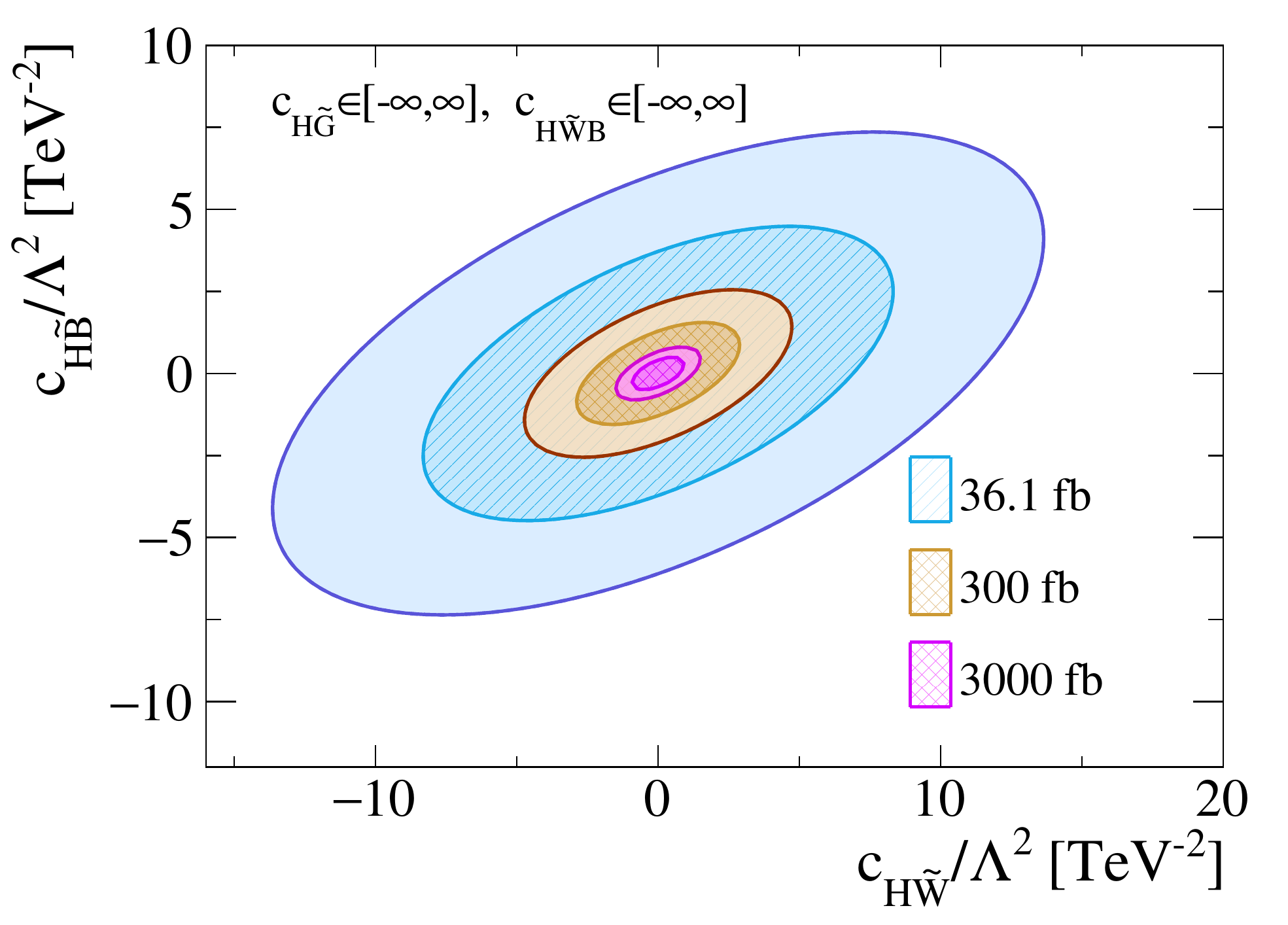}}
\caption{\label{fig:lumi_chgt_chwt_chbt_chwt} Dependence of the 2D constraints for CP-odd Higgs boson interactions on the integrated 
luminosity of the available dataset, without (panels (a) and (c)) and with (panels (b) and (d)) marginalisation over other CP-odd coefficients. All proposed measurements are included.  
The results are obtained using pseudo-data, and inner and outer shaded regions for each luminosity scenario represent the 68.3\% and 95.5\% CI, respectively.}
\end{figure*}

Although the blind directions can be lifted with the current dataset, the obtained constraints on CP-odd operators that affect the Higgs boson coupling 
to weak bosons are relatively weak ($c_i/\Lambda^2>1$~TeV$^{-2}$). This will be improved by increasing the integrated luminosity to increase the 
precision of these measurements. 
In Fig.~\ref{fig:lumi_chgt_chwt_chbt_chwt} and 
Table~\ref{tab:lumi}
we present the expected 1D and 2D constraints with larger datasets of 
300/fb (corresponding to the end of LHC Run-3) and 3000/fb (corresponding to the end of HL-LHC), for the full combination of 
differential measurements we consider.  Since the uncertainties are dominantly statistical, a simple extrapolation should be accurate up to the highest expected luminosities.  For the HL-LHC the results improve dramatically and the constrained values of $c_i/\Lambda^2$ approach unity.  To demonstrate the perturbative validity of the expected constraints, the magnitude of the interference contribution to the most sensitive distribution, relative to the SM contribution, is estimated using the left-hand side of Eq.~\eqref{eqn:eft_valid} and summarised in Table~\ref{tab:validity} for datasets of 300/fb and 3000/fb.

\begin{table}[!htpb]
\centering
\begin{tabular}{|c|c|c|c|}
\hline\hline
Coefficient  &  &  &  \\
$\left[\mathrm{TeV}^{-2}\right]$ & 36.1 fb$^{-1}$ & 300 fb$^{-1}$ & 3000 fb$^{-1}$ \\
\hline
\rule{0pt}{\normalbaselineskip} 
$c_{H\tilde{G}}/\Lambda^2$ & $[ -0.19 , 0.19 ]$ & $[ -0.067 , 0.067 ]$ & $[ -0.021 , 0.021 ]$ \\
$c_{H\tilde{W}}/\Lambda^2$ & $[ -11 , 11 ]$ & $[ -3.8 , 3.8 ]$ & $[ -1.2 , 1.2 ]$ \\
$c_{H\tilde{B}}/\Lambda^2$ & $[ -5.9 , 5.9 ]$ & $[ -2.1 , 2.1 ]$ & $[ -0.65 , 0.65 ]$ \\
$c_{H\tilde{W}B}/\Lambda^2$ & $[ -14 , 14 ]$ & $[ -4.9 , 4.9 ]$ & $[ -1.5 , 1.5 ]$ \\[0.5ex]
\hline \hline
\end{tabular}
\caption{Expected 1D constraints on Wilson coefficients for each EFT operator, in units of TeV$^{-2}$, after marginalising over all other coefficients.\label{tab:lumi}}
\end{table}%

\begin{table}[!t]
\centering
\begin{tabular}{|c|K{3.2cm}|K{3.2cm}|}
\hline\hline
Coefficient &  \multicolumn{2}{c|}{Allowed magnitude of CP-odd contribution} \\
 $\left[\mathrm{TeV}^{-2}\right]$ & 300 fb$^{-1}$  & 3000 fb$^{-1}$ \\
\hline
\rule{0pt}{\normalbaselineskip} 
$c_{H\tilde{G}}/\Lambda^2$  & $33\%$ & $10\%$ \\
$c_{H\tilde{W}}/\Lambda^2$  & $47\%$ & $15\%$ \\
$c_{H\tilde{B}}/\Lambda^2$  & $8\%$ & $2\%$ \\
$c_{H\tilde{W}B}/\Lambda^2$ & $25\%$ & $8\%$ \\[0.5ex]
\hline \hline
\end{tabular}
\caption{Expected sum of the moduli of the positive and negative interference contributions from CP-odd operators relative to the SM cross-section, see Eq.~\eqref{eqn:eft_valid},
allowed by the constraints in Table~\ref{tab:lumi} at a given luminosity.\label{tab:validity}}
\end{table}%

It is worth noting that the Run-3 and HL-LHC constraints presented above are simple extrapolations of current ATLAS results 
(and those that are already possible) to higher luminosities, and a number of other measurements can in principle be made that 
would tighten the constraints further. For example, all the constraints should trivially improve by about a factor of $\sqrt{2}$ if 
the proposed measurements are made by both ATLAS and CMS. In addition, as the datasets increase, splitting the measurement of 
the signed-$\Delta\phi_{jj}$ observable into VBF-enhanced and VBF-suppressed phase spaces will also be possible in the 
$h\rightarrow ZZ^\ast \to 4\ell$ decay channel.  Furthermore, model-independent $\Delta\phi_{jj}$ measurements in the $h\rightarrow WW^\ast\to\ell\nu\ell\nu$ and 
$h\rightarrow \tau\tau$ decay channels, as well as differential cross sections as a function of the decay angles in 
$h\rightarrow WW^\ast\rightarrow \ell\nu\ell\nu$ decay would add further constraints. Finally, model-independent differential measurements 
of other processes will be possible by the end of Run-3 and/or HL-LHC, with CP-sensitive differential information expected 
for Higgs boson production in association with a weak boson \cite{Godbole:2013lna} or a top-antitop quark pair \cite{Goncalves:2018agy,Buckley:2015vsa}. The measurements of Higgs 
boson production in association with a weak boson would add additional information that could constrain the $O_{H\tilde{W}}$, 
$O_{H\tilde{B}}$ and $O_{H\tilde{W}B}$ operators. Measurements of Higgs boson production in association with a top-antitop pair 
would constrain CP-violating complex phases in the EFT operators corresponding to the Yukawa sector, thus removing the blind 
direction between those operators and $O_{H\tilde{G}}$ that is implicit in this analysis.

\section{Conclusions}
\label{sec:conc}
A better understanding of the Higgs-boson properties remains a crucial part of the LHC phenomenology programme, offering a wealth of opportunities to connect the electroweak scale with other well-established features of beyond-the-SM physics. In this sense, the search for CP-violation in the Higgs sector is a crucial piece of the puzzle of the TeV scale.

In this Letter, we consider CP-violating operators in the context of gluon-fusion and vector-boson fusion production of Higgs bosons in association with jets. By focusing on the SMEFT approach, linearised in the Wilson coefficients, 
we can separate CP-odd and CP-even Higgs boson BSM interactions.
The former are then contained in asymmetries of genuinely CP-odd observables. 

We combine existing ATLAS differential measurements of the signed azimuthal angle between jets in $h+2~\rm{jet}$ production, and calculated a combined asymmetry in this angle of $0.3 \pm 0.2$.  An asymmetry of this magnitude would lead to the question of its origin, and we use its presence to discuss measurements that can be performed with existing data to further characterize the potential source.  In particular, separating the weak and strong production of the Higgs boson, and supplementing the current analyses with precision measurements of the CP-sensitive angle $\Phi$ for $h\rightarrow ZZ^* \rightarrow 4\ell$ decays, should break the degeneracies in the CP-odd coupling space. 
As our results are purely driven by asymmetries, it was not a priori clear that the LHC would be able to obtain perturbatively meaningful constraints with interference-only fits to dimension-6 Wilson coefficients. We show that although the current statistical uncertainties on the measurements are too large to provide constraints that are meaningful when compared to perturbative UV completions, LHC projections suggest that the Wilson coefficients will be constrained to unity or better for new-physics scales of $1~\text{TeV}$.

\acknowledgements
We thank Tilman Plehn for helpful discussions. \\
F.\ B.\ is supported by the DFG Emmy-Noether Grant No.\ BE 6075/1-1 and thanks the Aspen Center of Physics, supported by
the NSF grant PHY-1066293, where part of this Letter was completed. C.\ E.\ is supported by the IPPP Associateship scheme and by the UK Science and Technology Facilities Council (STFC) under grant ST/P000746/1.
C.\ H.\ is supported by the IPPP Associateship scheme.
K.\ L.\ and H.\ M.\ are supported by the European Union's Horizon~2020 research and innovation programme under ERC grant agreement No.\ 715871.
A.\ P.\ is supported by the Royal Society under grant UF160396 and by an IPPP Senior Experimental Fellowship.
D.\ P.\ is supported by STFC under grant ST/M005437/1 and by an IPPP Senior Experimental Fellowship. 

\newpage
%

\end{document}